# Vibe coding for clinicians: democratising bespoke software development for digital health innovation

Ariel Yuhan Ong,[1,2,3,4] Iain Livingstone,[5] Caroline Kilduff,[2] Mertcan Sevgi,[1,2,3] David A Merle,[1,2,3] Eden Ruffell,[1,3] Pearse A Keane,[1,2,3] Fares Antaki[6,7]

1. Institute of Ophthalmology, University College London, UK
2. Moorfields Eye Hospital, Moorfields Eye Hospital NHS Foundation Trust, London, UK
3. NIHR Moorfields Biomedical Research Centre, London, UK
4. Oxford Eye Hospital, Oxford University Hospitals NHS Foundation Trust, Oxford, UK
5. NHS Forth Valley, Scotland, UK
6. Cole Eye Institute, Cleveland Clinic, Cleveland, OH, USA
7. The CHUM School of Artificial Intelligence in Healthcare, Montreal, QC, Canada

Correspondence to Ariel Yuhan Ong ariel.ong@nhs.net and Fares Antaki fares.antaki@outlook.com

## ABSTRACT

Clinicians often face workflow problems that are perceived as either too bespoke or low stakes to attract commercial attention. Historically, most do not have the technical knowledge to address these problems, but the recent emergence of “vibe coding” presents a transformative opportunity. Vibe coding refers to the co-development of software using natural language prompts to large language models. It offers a pathway to create simple tools that address these real-world pain points, or to prototype more complex ideas. In this review, written by a group of early adopter clinicians with a range of programming expertise, we introduce vibe coding for clinicians (especially those with no or minimal coding experience) as a way of democratising innovation from the front lines. We discuss foundational skills, outline some common challenges, provide a practical step-by-step playbook, and illustrate this approach with some case examples. We also propose a risk stratification framework which links the target audience and intended use of developed tools to a corresponding set of development and governance expectations. Vibe coding can bridge the gap between clinical insight and technical execution, equipping clinicians with the ability to rapidly prototype digital health solutions most reflective of clinical realities, but caveats and guardrails for safe deployment must be considered in order to optimise patient benefit.

Modern medicine is plagued by a “long tail” of software needs – problems in clinical workflows, research processes, and medical education that may be perceived as too niche or low-stakes to justify commercial development or institutional investment upfront without further work. Large-scale solutions such as electronic health record (EHR) systems address broad operational requirements but often lack regard for usability or the consideration towards the granularity required to solve hyper-specific, context-dependent challenges.[1] There is a strong case for inclusive co-creation in developing digital health solutions.[2] Instead, these issues are compounded by a persistent misalignment between software developers and clinical realities, which limits the full potential of health technologies in clinical practice.[3,4] Collectively, these challenges impose a significant burden on the efficiency of clinical workflows, and contribute to clinician stress and burnout.[1]

While clinicians possess the domain expertise necessary to identify and articulate these needs, most lack the programming skills to translate their insights into functional solutions, even as access to financial resources to engage a professional software developer for small-scale projects may be limited. As a result, tasks with high clinical value but no direct pathway to reimbursement are often neglected. For example, a simple web application could generate a personalised postoperative schedule for patients who have been prescribed multiple medications, mitigating the risks of non-adherence and downstream complications. Another valuable tool might be a simple risk calculator designed to enhance the shared decision-making and consent process for a specific procedure.

**A brief introduction to vibe coding**

Recent advancements in artificial intelligence (AI) offer a potential solution to bridge this gap. The generative abilities of large language models (LLMs) have led to the emergence of “vibe coding”, an intuitive, iterative, and conversational process wherein individuals use natural language prompts to co-develop functional software with an LLM or LLM-based tool, without having to write precise and syntactically flawless code.[5,6] This term, coined in early 2025, reflects a spectrum of engagement from complete reliance on the LLM to responsible AI-assisted development or collaboration.[7] Vibe coding builds on the trajectory of low-code/no-code platforms,[8] and represents the next step in the evolution of programming assistance – a journey that began with community forums for crowdsourcing tips and advice (e.g. Stack Overflow[9]), and AI-powered autocompletion tools (e.g. GitHub Copilot[10]).

While much of the discourse on vibe coding exists as public accounts on social media platforms, formal analyses of its practical applications, risks, and benefits are beginning to emerge, latterly in the medical literature. Recent work analysing expert programmers’ vibe coding workflows found that rather than eliminating the need for programming expertise, this approach enabled them to focus on context management and direction.[11] For users without programming experience, early work has demonstrated the benefits of this no/ low-code development approach in allowing medical educators to create interactive simulations.[12,13] Work led by our group in clinicians with little to no coding experience demonstrated early promise in their ability to develop digital health apps, although with important caveats.

In this review, we aim to provide a framework for non-technical clinicians to engage with vibe coding effectively and responsibly, informed by our experiences as early adopter clinicians with a range of programming expertise. We also aim to highlight the foundational skills and common challenges inherent in this process, and to discuss the role of vibe coding within a broader systems perspective.

**Vibe coding for clinician-led development**

The emergence of vibe coding is perhaps analogous to the personal computer (PC) revolution of the late 20$^{th}$ century. When PCs entered individuals' homes, the everyman, who was a domain expert in their own life and work, was empowered to build solutions for problems that major tech companies had never considered. For example, spreadsheets for family budgets and custom databases for small businesses flourished because the people most familiar with the problems had access to the tools to start building the answers. In the same vein, 3D printing has also revolutionised physical creation across multiple industries by facilitating rapid prototyping and customisation.[14] This has led to the development of personalised patient-specific anatomical models for preoperative planning, customised implants and drug delivery devices, as well as surgical and research tools.[15,16] Vibe coding presents a similar paradigm shift for healthcare, by placing powerful development tools in the hands of clinicians who are experts in the complex domain of patient care.

For clinicians, this approach offers several key benefits. It democratises access to software development by empowering them to build a bespoke minimum viable product (MVP) that addresses their pain points directly. Beyond simple storyboards or sketches, the act of building and testing an executable artefact allows clinicians to reason through design by doing. Functional MVPs - even rudimentary ones - can expose workflow misalignments, usability issues, and hidden dependencies that static mock-ups cannot show, while facilitating user testing and feedback. This affords clinicians greater autonomy and enables rapid iteration at low cost. Demonstrating a potentially viable solution provides early evidence of value by evidencing that a seemingly niche or low-stakes problem is worth solving. This form of pragmatic "innovation triage" also helps to ensure that only validated and well-defined ideas progress further, for example to formal development, which is particularly valuable in settings with constraints on financial and technical resources.

**Vibe coding: a step-by-step guide**

Vibe coding can be conducted in three main ways **(Table 1)**.

The most direct method involves interacting with a general-purpose conversational LLM (e.g. ChatGPT, Gemini, Claude) as a programming partner and receiving blocks of code to be run in a separate integrated development environment (IDE) or a simple text editor to write, organise, and run the code. However, this workflow is inherently fragmented, and requires repeated context switching, manual code integration, and explicit user oversight for execution, debugging, and state management.

Alternatively, a more beginner-friendly approach involves using integrated cloud-based LLM-powered platforms (e.g. Lovable, Replit) which can be run in the browser, and which streamlines the process by combining a code editor, a conversational AI, and a live execution environment. These systems reduce friction by bundling development processes into a single streamlined workflow, as well as building a full stack application - both the front end (the parts of a website or application that users interact with) and the back end (the server-side logic, databases, and application infrastructure). Some also support multimodal input such as voice interaction, allowing users to describe an application verbally.

Finally, agentic coding workflows represent a more advanced approach where the AI acts as a goal-directed agent, autonomously planning, executing, and iterating on multi-step development tasks. In this paradigm, the user's role becomes that of a systems architect and strategic planner. While these workflows do not necessarily require coding expertise, they are often most effectively leveraged by experienced developers managing multiple complex workflows and large code bases. While some consider agentic coding separate from vibe coding,[17] it can be conceptualised as an assistive subset, particularly when human-in-the-loop checkpoints are in place (e.g. Claude Code).

**Table 1: A comparison of the different tools available for vibe coding - chatbots, browser-based integrated development environments (IDEs), and agentic coding workflows - and their functionalities, use cases, and safety considerations.**

| | General purpose LLM/ chatbots | Integrated platform | Agentic coding workflows |
|---|---|---|---|
| **Examples** | OpenAI - GPT-5 series; Anthropic - Claude 3 (Opus, Sonnet, Haiku); Google DeepMind - Gemini 2.5 (Pro, Flash)<br>Note, Claude offers the “artifacts function” that is more similar to an integrated browser-based platform) | CodeSandbox, Lovable, Replit, Cursor, Google Antigravity | Claude Code, Devin, Open Claw |
| **Description** | Conversational AI models that generate, explain, and refactor code based on user prompts. Can be either cloud-based (proprietary models) or run locally (local models) | All-in-one IDEs that combine a code editor, conversational AI, and a live execution environment. May be cloud- or browser-based, or local | Autonomous or semi-autonomous systems that plan and execute complex, multi-step coding tasks |
| **Potential use case(s)** | Exploration, generating snippets, learning new concepts, and debugging isolated code blocks. | Prototyping, building full-stack MVPs, and collaborative projects | Professional developers managing multiple complex workflows and large |

| | | | code bases |
|---|---|---|---|
| **Safety** | User must validate all code for security flaws and code correctness | Provides options for validating safety mechanisms, depending on the platform | Requires consideration of sandboxed execution due to the risk of unintended actions from complete autonomy |
| **Learning curve** | Low to moderate: intuitive user interface, but user needs to learn effective prompt engineering and appropriate IDEs to deploy the generated code | Low: best suited to beginners, end-to-end conceptualisation to deployment | Moderate to high, requires software engineering knowledge for the best results |
| **Costs** | Free for local models. Tiered subscription model for proprietary models | Typically a tiered subscription model | Typically more expensive subscription models or based on token consumption |
| **Strengths** | Speed and versatility for focused tasks | Low barrier to entry<br>Streamlines the end-to-end workflow, including deployment, hosting, version control integration, database setup and integration. | Automates complex and/or tedious tasks |
| **Limitations** | Risk of insecure code generation and maintainability as general LLMs do not innately consider these<br>Hallucinations<br>Lack of knowledge of full code base (if there are context window limitations) | Vendor lock-in; entire code base lives on their online platform<br>Constrained by platform user caps | Limited transparency for agent's actions<br>Can be complex to set up and control |

LLM, large language model; MVP, minimum viable product

The process of vibe coding can best be described as a structured multi-stage cycle. In our experience, the core workflow typically comprises the following key elements **(Figure 1)**:

1. **Defining the vision and scope.** This is the most fundamental step of the entire endeavour. It involves defining the functionality and aesthetics of the desired tool, and sketching out the MVP (the simplest version which performs the most essential function). Although having a well-defined blueprint is not strictly necessary at this stage, a clearly defined goal helps guide the development cycle, reducing the risk of ambiguous instructions and frustrating iteration loops.

2. **Translating the vision into an effective prompt.** An effective initial prompt should provide as much context as possible. This may include specifying the technology stack (e.g. HTML, CSS, JavaScript), which may be less necessary for integrated browser-based platforms; describing the required features from the initial outline; and clearly articulating the core functionality of the MVP. Essentially, better prompting can improve code quality. Using an LLM to help articulate these abstract concepts can sometimes be helpful.

3. **Testing** (also known as “integration testing” in software development terms). The code generated by the LLM is then tested in an IDE of the user’s choice, for example by copying this over into a local application, or more seamlessly on browser-based platforms where the generated code can be run instantly. Beyond testing functionality, looking for errors and learning to read error messages becomes essential here. A second step is checking whether the code remains reliable across different scenarios. Introducing simple edge cases can help reveal hidden assumptions and ensure that the solution holds up more broadly.

4. **Debugging and iterative refinement** is the core feature of vibe coding. While simply copying and pasting the error message is the most straightforward and low-effort approach, this process becomes an invaluable learning opportunity when the user starts seeking explanations, transforming an often-frustrating step into an opportunity for interactive tutoring. Prompting the LLM to understand whether the solution produced is necessarily the most efficient and maintainable can be helpful here as well.

5. **Documenting and versioning** are good software development practices essential for developing a reproducible and maintainable project. In practice, these considerations should begin at the ideation stage. Documentation involves saving the sequence of key prompts and adding (or instructing the LLM to add) comments within the code to explain their purpose. Versioning involves saving snapshots of the project at key milestones, either as simple file copies or using more robust version control systems such as Git.

6. **Repeating the cycle.** The above steps form a continuous iterative loop (akin to an “agile software development cycle”). Once a feature has been successfully tested, implemented, and documented, the cycle begins anew for the next desired feature or refinement. Each additional feature should be added iteratively through small, specific, and incremental prompts instead of vague, sweeping commands which risk producing unwanted changes elsewhere, in order to ensure a controlled and stable development process.

**Figure 1: The vibe coding workflow illustrating the iterative cycle of software development with a large language model (LLM).** The core loop begins with the clinician 1) defining the vision for the project, which is then translated into a 2) natural language prompt. The LLM generates code, which is then 3) tested in a suitable environment to observe its functionality and identify errors and areas for improvement. Any errors or desired changes are addressed in the 4) iterative refinement and debugging stage through conversational feedback to the LLM. This loop is then repeated, allowing for the incremental addition of features and refinement of the application. This workflow is underpinned by 5) the documentation of key prompts and comments for clarity, as well as versioning the code to track progress. The cycle is then repeated.

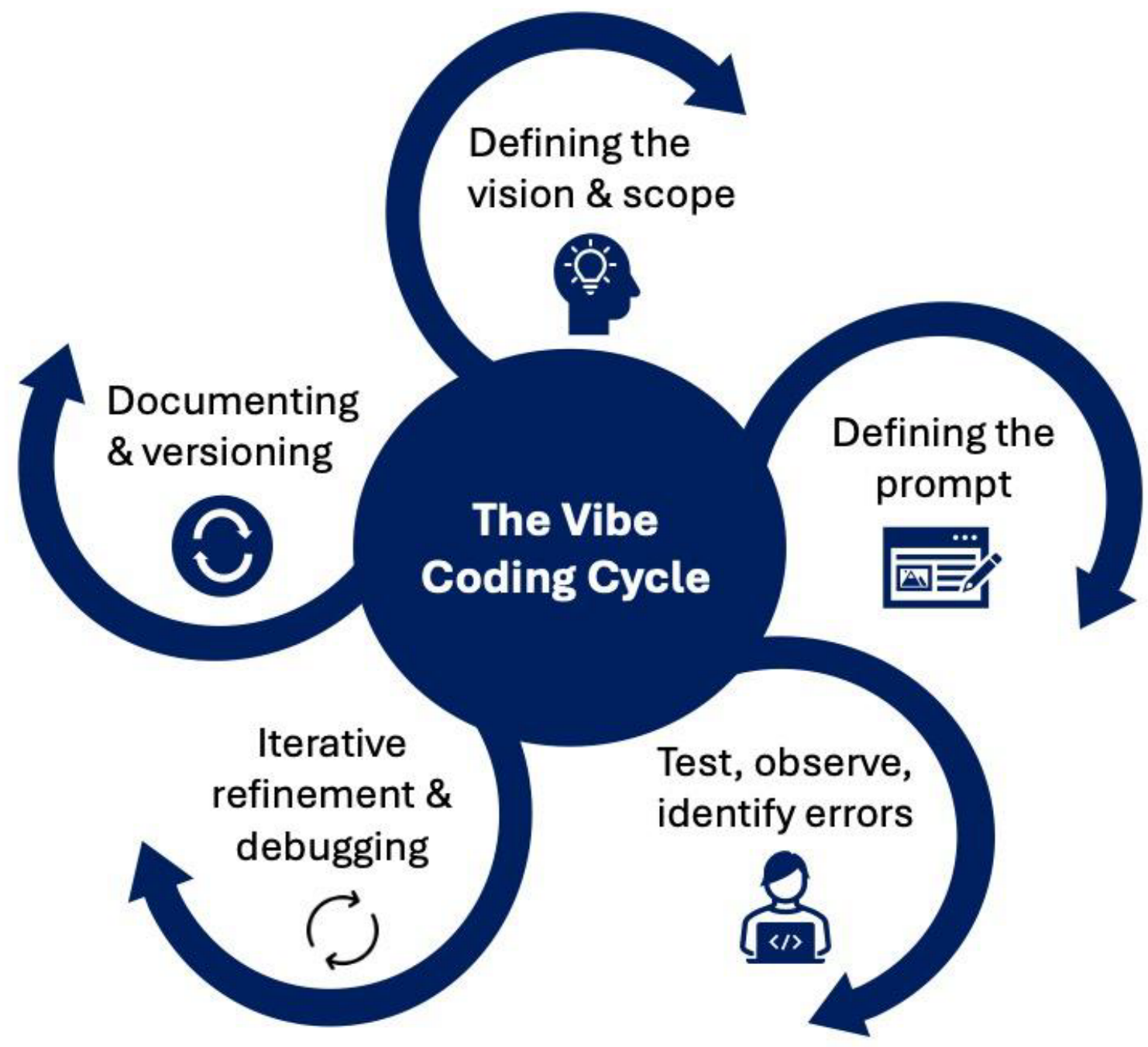

While a scattershot trial-and-error approach to vibe coding is possible, for clinicians without coding experience, adhering to these general principles helps to minimise the risk of common pitfalls and reduce user frustration. This also helps to ensure that the code base for the final prototype is reproducible and maintainable, as well as being ready for collaboration, for example in “pair programming” with experienced developers to further refine the prototype.

A grounding in the basics of coding and core software development principles is perhaps the most essential requirement to amplify the benefits of LLM-assisted development and minimise the accrual of technical debt. This includes appreciating when a programming language or paradigm is a poor fit for a particular task. For example, developing an application in pure Python may not be the most effective approach. Having a working understanding of object-oriented, functional, and procedural styles helps the user to judge whether the structure of the generated code aligns with the use case. Understanding how “state” behaves in the context of the programme is also essential for tracking how variables evolve, propagate, or are shared across functions and classes. Current LLMs do not inherently optimise for long-term project maintainability, and can readily generate convoluted “spaghetti code”. Without understanding this, beginners risk falling victim to the Dunning-Kruger effect[18] – feeling confident in their creation while unknowingly accumulating significant technical debt that renders the prototype unmaintainable or difficult to build upon.

Key principles to be aware of include: 1) version control, or the practice of managing changes to code; 2) reproducibility, or ensuring that the development process can be reproduced by saving the sequence of key prompts alongside the code; 3) clear documentation, which involves the addition of comments within the code to explain the function of each code block for future reference; and 4) maintainability, or writing clean, logical, and structured code that is easy to modify or debug in the future. Beginners can promote maintainability by instructing the LLM to break down the application into logical components and to use clear naming conventions. Treating documentation and versioning as habits rather than retrospective obligations is perhaps the single most important shift in practice for clinicians new to software development.

Finally, foundational skills such as basic proficiency in user interface/ user experience (UI/UX) design principles and basic AI literacy are essential. Understanding UI/UX principles will help in ensuring that the software is usable - this may include intuitive navigability, legible typography, strategic placement of primary action buttons, clear presentation of error states, and effective communication of task progression. At the same time, AI literacy is necessary for users to fully appreciate that LLMs are probabilistic text prediction models and to minimise their tendency towards anthropomorphisation.[19] This is useful for contextualising LLM outputs and encouraging a healthy skepticism to reduce the risk of overreliance, which is essential given the risks of hallucinated code[20] in the context of an asymmetry in domain expertise between users and LLMs. In addition, while specific prompting techniques may evolve with advances in LLMs, understanding the basics of prompt engineering and the need to craft clear and context-rich instructions to provide effective guidance remains essential for optimising the quality of the outputs. This may even involve a meta-level request asking the LLM (or a separate LLM) to optimise the prompt for clarity and detail).[21]

**A risk stratification framework for vibe coded tools**

However, this newfound capability is not without risks, including the potential proliferation of low-quality, unmaintainable or insecure software. We posit that while vibe coding may produce a useful prototype, it should not be used by itself to create high stakes regulated medical devices or any function where failure could lead to patient harm without appropriate technical knowledge or input. To support clinicians considering this, we propose a pragmatic three-tier classification which links the target audience and intended use to a corresponding set of development and governance expectations **(Table 2)**:

- Tier 1 includes tools intended for personal use, which do not involve patient data nor have any bearing on clinical decision-making. At this tier, the principal safeguard is the developer’s own testing and judgment, supplemented by basic documentation of the prompts and logic used.

- Tier 2 encompasses administrative or workflow tools used within a team or department which may indirectly shape clinical workflow (e.g. scheduling tool). The logic of the tool should be reviewed by a second person; and a named individual should be responsible for maintenance. The relevant IT or information governance team should be notified, particularly where the tool interfaces with any hospital network or infrastructure.

- Tier 3 encompasses tools that are patient-facing, handle identifiable data, or have the potential to influence clinical decisions directly. At this tier, formal validation and institutional governance review is required prior to any real-world deployment. Professional software development input is strongly recommended. Tools at this tier may qualify as software as a medical device (SaMD) under applicable regulatory frameworks, and should be assessed accordingly.

This classification is intended to ensure that the ambition of a project is matched by an appropriate level of due diligence, rather than discouraging development at higher tiers. In practice, there are several potential outcomes for these vibe coded prototypes. A personal prototype can transition into a Tier 2 or Tier 3 tool as its scope or user base expands, and the corresponding safeguards should scale accordingly. Some may become features integrated into larger commercial platforms such as EHR modules through close collaboration with vendors' design authority. Others may mature into standalone commercialisable products, of which a small subset may require regulatory approval depending on the use case. Many may remain as personal tools, local administrative workflow solutions that improve the efficiency of daily tasks, educational tools that improve training, or research tools that facilitate data collection or academic collaboration. Collectively, these pathways illustrate that the value of vibe coding lies in supporting clinicians to develop prototypes and safely explore ideas to address unmet needs, identify those with a genuine potential or need to attract support and funding, and then take these forward for technical implementation as needed.

**Table 2: A proposed risk stratification framework for vibe coded tools developed for healthcare or healthcare-adjacent purposes.**

| | Tier 1: Personal tool | Tier 2: Administrative or workflow tool for team or department | Tier 3: Patient-facing or clinical decision support |
|---|---|---|---|
| **Influence on clinical decisions** | None | Indirect influence | Direct influence |
| **Regulatory considerations** | N/A | Likely N/A but will need to reassess if scope expands | Medical device regulations will likely apply |
| **Validation** | Testing by developer | Testing by developer +/- professional developer | Institutional review required; professional developer input required |
| **Governance** | N/A | May require IT or information governance input; developer input likely beneficial Requires named person to be responsible for maintenance | Information governance review required; professional developer input highly recommended |
| **Documentation** | Prompt log, code comments, version control | Prompt log, code comments, version control | Full SDLC documentation Software bill of materials (SBOM) |
| **Example** | Date or interval calculator for personal use<br>Medical education tool for personal use | Department scheduling tool<br>Image segmentation tool | Clinical risk calculator<br>Patient-facing chatbot |

AI, artificial intelligence, N/A, not applicable

**Case examples**

The following case studies illustrate how clinicians with varying degrees of coding experience have used vibe coding to develop practical solutions for real-world problems.

***Case 1:*** Intravitreal injections are a mainstay of ophthalmology services and are among the most ubiquitous procedures performed in clinical practice.[22] Treatment typically follows a “treat-and-extend” regimen (in 2-4 week increments or decrements from 4-16 weeks), where clinicians titrate the treatment interval based on the current clinical picture.[23] However, appointments may be frequently delayed in a busy service, or patients may miss their appointments for various reasons, meaning that calculating the treatment interval from the last injection to determine the next becomes a tedious task which few EHRs have considered automating. Manually counting weeks on a calendar is inefficient and prone to error, especially under time pressure in a high-volume clinic, and the few online date calculators lack the necessary functionality or task efficiency.

To address this, a simple web application was vibe coded using a browser-based platform **(Figure 2)**. A series of plain language prompts described the desired tool: a clean professional interface with two date selection fields, with the latter designed to default to the current date for ease of use while remaining editable. The application was instructed to calculate the time interval between the two dates and display the result clearly in both weeks and days. Through several rapid iterations, the final functional prototype was completed in less than fifteen minutes, saving valuable clinician time (~30 seconds per patient, which compounds over the course of multiple high-volume clinics). Under the risk stratification framework proposed above, this tool is used by its developer alone and performs a simple arithmetic function, and as such, falls within the Tier 1 category. basic quality control was therefore undertaken as part of good practice, and the underlying code and outputs were manually reviewed to confirm accurate interval calculations and consistent behaviour before use. Its value lies not in technical sophistication but in the elimination of a small, persistent source of cognitive load across a high-volume workflow – an archetype of the types of problems that vibe coding is particularly well suited to address.

**Figure 2: The user interface for a simple intravitreal injection interval calculator featuring two date selection fields, with the latter designed to default to the current date for ease of use while remaining editable.** The application was instructed to calculate the time interval between the two dates and display the result clearly in both weeks and days.

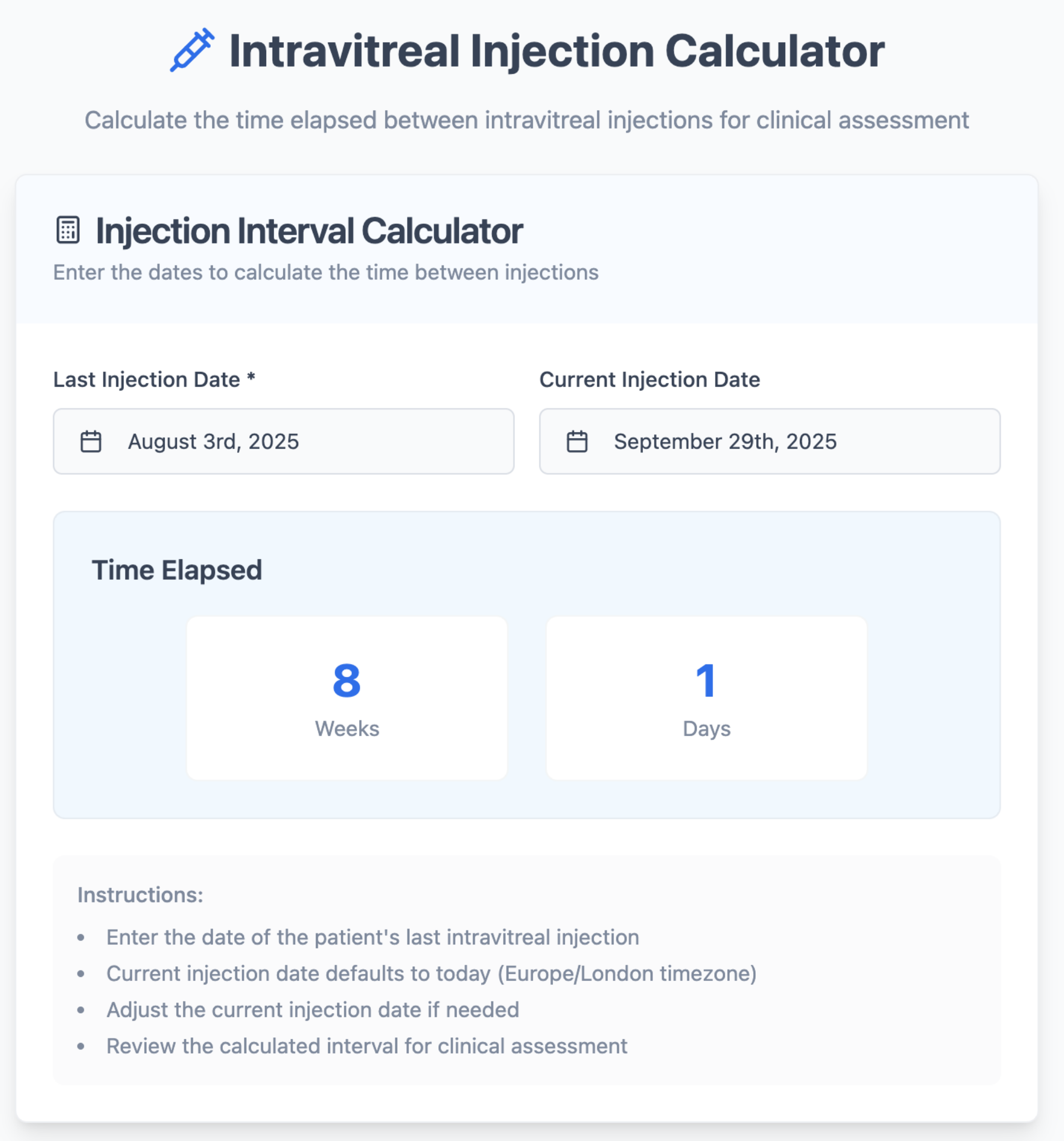

***Case 2:*** Postoperative treatment regimens are designed to improve healing and surgical outcomes. However, non-adherence following ophthalmic surgery can be substantial – electronic compliance monitoring studies report a mean dose compliance of 50%, and premature treatment discontinuation is common, often owing to forgetfulness and confusion around complex treatment regiments.[24,25] Administration is complicated by surgeons prescribing one to three eyedrops postoperatively, with variable dosing intervals and sometimes a tapering regimen that further complicates adherence.[26]

In 2024, the Moorfields Launchpad, a hospital-based forum for innovation, designated postoperative eyedrop adherence as a key challenge for their hackathon. The winning solution ("Cataract Calendar") provides a personalised dosing calendar to support patients in managing their treatment regimen. This was designed as an online tool which can be downloaded as a PDF **(Figure 3)**. Laterality, surgery date and diabetic status is entered; the system then generates a month-by-month schedule showing the type, number and timing of drops per day. This was designed to be printable in black-and-white (suitable for NHS workflows) for patients who prefer a paper-based schedule. The prototype was vibe coded using Lovable (a low-code browser-based AI-powered platform designed for the express purpose of facilitating vibe coding), and is being migrated to a subscription-based no-code platform.

In the context of the proposed framework, the Cataract Calendar falls within Tier 2, with a clear trajectory towards Tier 3 as it moves into patient-facing deployment within surgical workflows. As such, due diligence with quality control steps is planned prior to pilot testing in real-world surgical workflows on an opt-in basis, as is institutional information governance review. Formative feedback has been gathered from patients, and usability and impact on adherence will be validated prospectively. The tool will potentially be adapted for other long-term treatment regimens outside of postoperative care, as well as non-ophthalmic postoperative workflows beyond ophthalmology. This illustrates a point of broader relevance – the transition between tiers is not a barrier to progress, but rather serves as a structured checkpoint that ensures that a project's maturation is accompanied by the rigour its expanding scope demands.

**Figure 3: The “Cataract Calendar” tool featuring the online user interface which can generate a PDF version to be printed for patients who prefer a paper-based calendar.** Different postoperative treatment regimens (in terms of drugs, frequency of treatment, and tapering regimens) can be pre-selected and customised for individual patients.

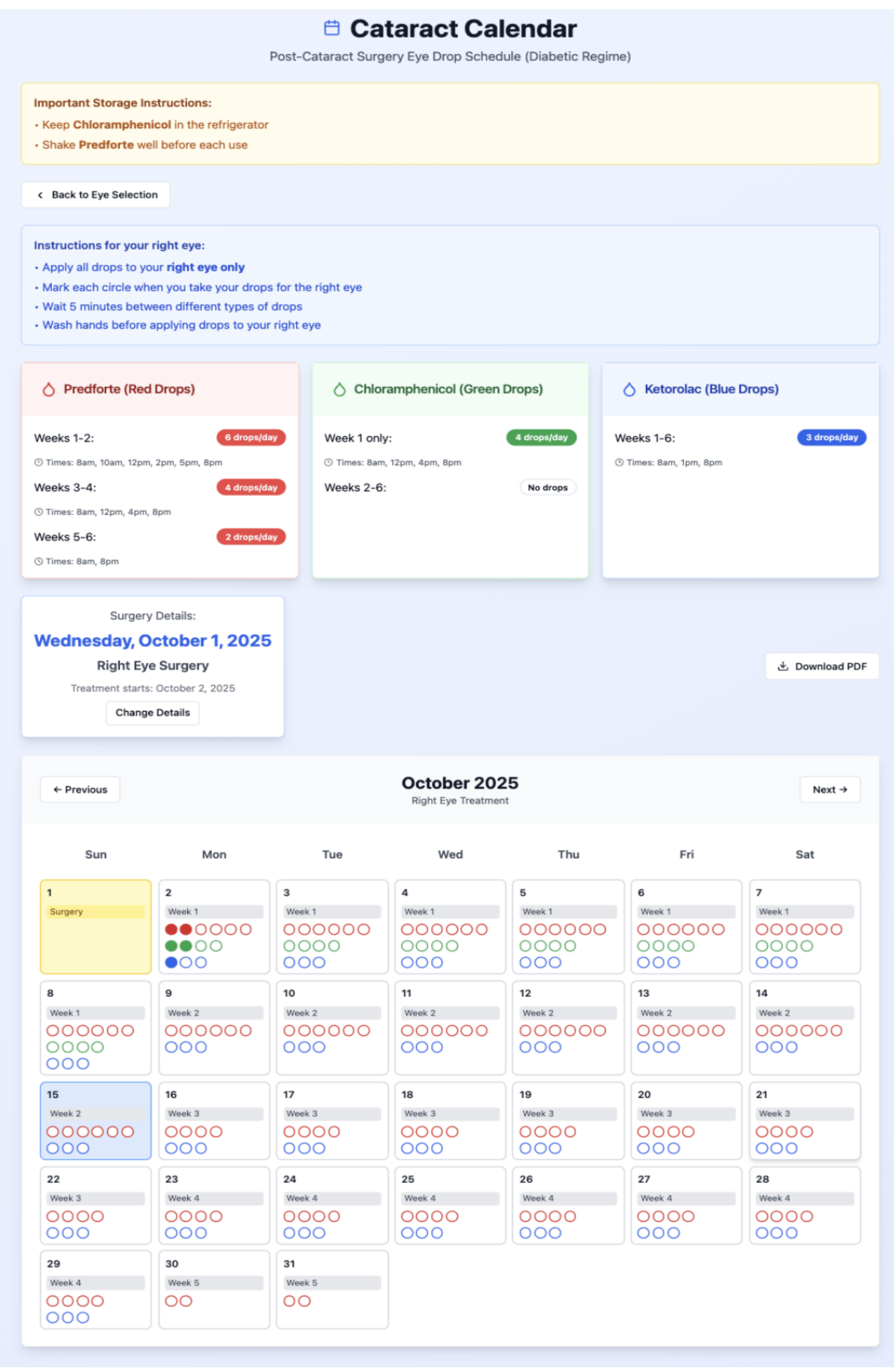

**Positioning vibe coding within the broader software development ecosystem**

The democratisation of software development does not equate to advocating for the rampant proliferation of poorly constructed applications. Instead, the value of vibe coding must be contextualised within the broader software development ecosystem, which requires an understanding of where informal experimentation ends and formal development begins. Not all vibe-coded software needs to be commercialised or widely deployed; many will remain valuable as “passion projects” for personal exploration, education, or local workflow support to solve a minor pain point for an individual or a small team.

While a tool intended for personal use carries limited institutional exposure, sharing it across teams or applying it to patient data introduces risks around security, long-term maintenance, and possibly even regulation. A related but often underappreciated risk is “shadow IT” – the informal deployment of tools outside institutional oversight.[32] Clinicians developing tools beyond Tier 1 prototypes should therefore engage proactively with IT and information governance teams to ensure appropriate security, support, and documentation. In the UK, frameworks such as the NHS Digital Technology Assessment Criteria (DTAC) may also provide a structured starting point for de-risking innovation and supporting safe deployment.[33]

The expectations change further for any project that shows promise for broader commercial use, redistribution, and/or clinical use as a medical device. This last should align with local regulatory guidance. For example, according to guidance from the Medicines and Healthcare Regulatory Agency (MHRA) in the UK, software qualifies as a medical device if it is intended for diagnosis, prevention, monitoring, treatment, or alleviation of disease or injury.[27] In all cases, provenance, authorship, and licensing must also be transparent due to the risk that vibe coding may pose to intellectual property (IP) rights. Empirical analyses of code-generating models have shown that a small proportion of outputs (0.88-2.01%) contain code fragments with striking similarities to existing open-source material, often without correct license metadata, especially for copyleft licenses.[28] Shipping unvetted AI-generated code into production therefore carries a non-trivial risk of distribution without attribution and breach of IP rights,[29] which is amplified in healthcare contexts by compliance and reputational stakes.[30]

Practical safeguards against this could include clear documentation of the LLM, version and prompts used; running basic licence-compliance or software composition scans, and rewriting any code of unclear provenance or code which demonstrate similarities; maintaining a lightweight software bill of materials (SBOM) to meet compliance and transparency requirements. Ultimately, alignment with regulatory standards and institutional assurance frameworks is essential for any application that moves beyond a personal prototype to involve patient data or influence clinical care,[31] in order to democratise innovation in a responsible manner while maintaining professional and legal standards.

This brings us to the most important caveat: just as AI in healthcare should augment rather than replace clinicians,[34,35] vibe coding does not replace professional software engineers or UI/UX designers, especially for vibe coded tools that transition into the formal software development life cycle. There is profound value in the act of building. When clinicians build their own prototypes, they are forced to translate abstract ideas into concrete logic and user flows. This process draws on their rich clinical expertise and deepens their own understanding, allowing them to articulate their vision in ways that accelerate and enrich collaboration with technical colleagues, as a working prototype is a far more powerful communication tool than a text document or a slide deck. The prototype can then serve as the starting point for a formal development process with technical experts.

## Conclusion

Vibe coding reframes clinician involvement in digital health from passive observation to active solution-oriented prototyping. The value proposition is threefold: inclusive co-creation leverages domain knowledge and expertise to produce user-centered prototypes that meet real-world needs; producing MVPs can highlight usability needs and data dependencies before a costly development process; and replacing abstract feature requests with working exemplars can reduce translation error and shorten iteration cycles, thus enriching collaboration with technical teams. Realising this potential requires certain guardrails, including a need for AI literacy, a basic understanding of software design principles, and structured handover into product engineering should commercialisation become a possibility. Used within these limits, vibe coding can help catalyse clinician-led innovation from the frontlines while preserving rigour and accountability.

## Author Contributions

AYO and FA conceptualised this manuscript. AYO wrote the first draft of the manuscript. AYO and CK prepared the figures. All authors (AYO, IL, CK, MS, DAM, ER, PAK, FA) critically revised and approved this manuscript for submission.

## Funding

AYO is supported by a National Institute for Health Research (NIHR) Doctoral Fellowship (NIHR303691). PAK is supported by a UK Research & Innovation Future Leaders Fellowship (MR/T019050/1), Moorfields Eye Charity with The Rubin Foundation Charitable Trust (GR001753), and an Alcon Research Institute Senior Investigator Award. The views expressed in this publication are those of the authors and not necessarily those of the abovementioned funding bodies.

## Competing interests

FA is an equity owner in SIMA Surgical Intelligence Inc. PAK is a cofounder of Cascader Ltd. and has acted as a consultant for Retina Consultants of America, Roche, Boehringer Ingelheim, and Bitfount, and is an equity owner in Big Picture Medical. He has received speaker fees from Zeiss, Thea, Apellis, and Roche. He has received travel support from Bayer and Roche. He has attended advisory boards for Topcon, Bayer, Boehringer Ingelheim, and Roche. None of the other authors report any conflicts of interest. None of the other authors report any conflicts of interest.